# Searching for Better Plasmonic Materials

*Paul R. West[1,+], Satoshi Ishii[1,+], Gururaj Naik[1,+], Naresh Emani[1], Vladimir M. Shalaev[1], and Alexandra Boltasseva [1,2,*]*

[1]School of Electrical and Computer Engineering and Birck Nanotechnology Center, Purdue University, West Lafayette, Indiana 47907, USA

[2]DTU Fotonik, Technical University of Denmark, Lyngby, DK-2800, Denmark

*Corresponding author: e-mail: aeb@purdue.edu

Telephone: 765-494-0301, Fax: (765) 494-6951

[+]These authors contributed equally to this work.

**Abstract**

Plasmonics is a research area merging the fields of optics and nanoelectronics by confining light with relatively large free-space wavelength to the nanometer scale - thereby enabling a family of novel devices. Current plasmonic devices at telecommunication and optical frequencies face significant challenges due to losses encountered in the constituent plasmonic materials. These large losses seriously limit the practicality of these metals for many novel applications. This paper provides an overview of alternative plasmonic materials along with motivation for each material choice and important aspects of fabrication. A comparative study of various materials including metals, metal alloys and heavily doped semiconductors is presented. The performance of each material is evaluated based on quality factors defined for each class of plasmonic devices. Most importantly, this paper outlines an approach for realizing optimal plasmonic material properties for specific frequencies and applications, thereby providing a reference for those searching for better plasmonic materials.

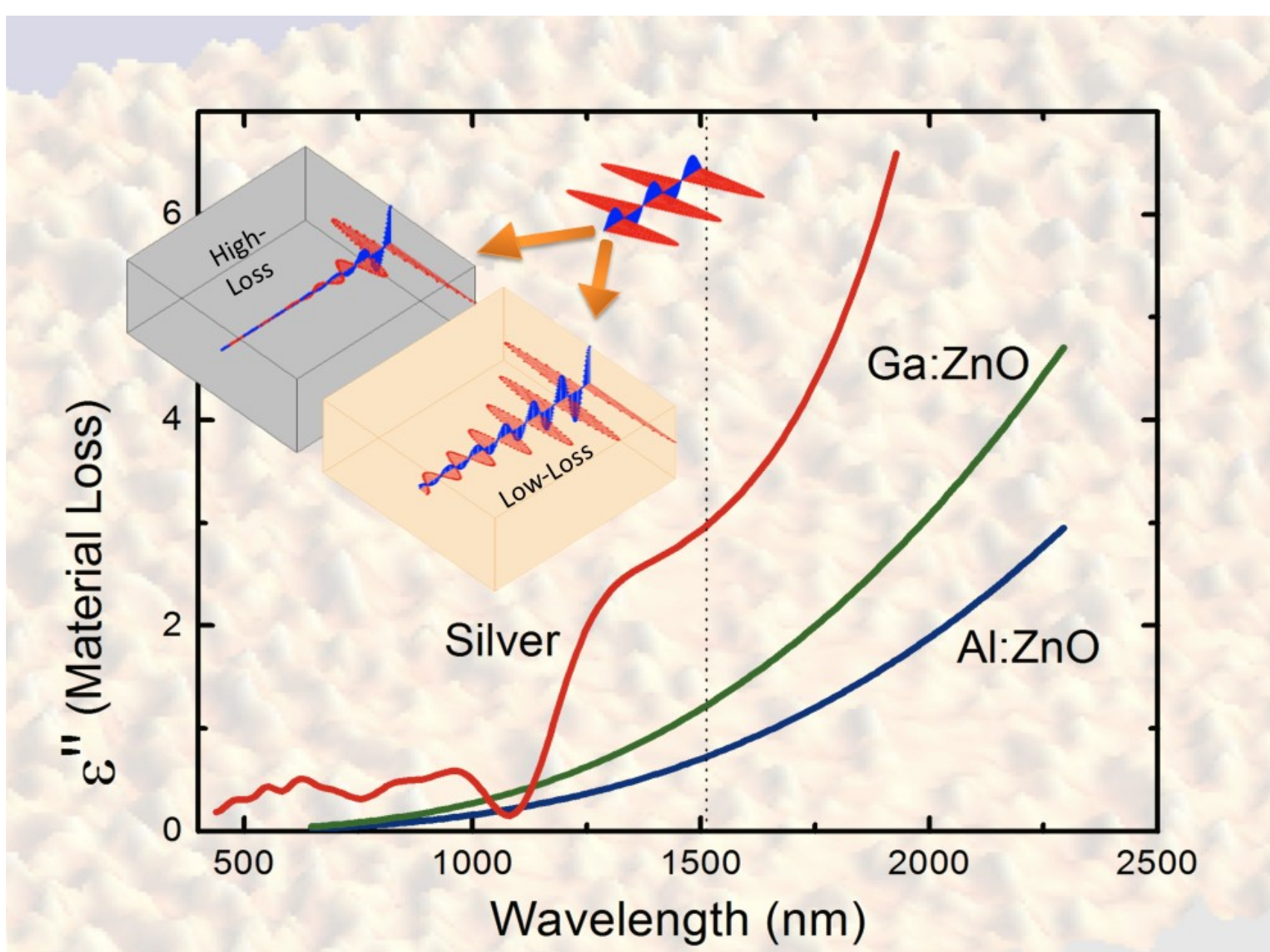

## 1. Introduction

The speed of information processing has seen rapid growth in the past few decades from the progress in scaling down the sizes of devices in micro- and nanoelectronics. However, researchers are experiencing major difficulties in reaching speeds over a few tens of GHz using this scaling approach due to fundamental limitations from RC-delays and power dissipation in the devices. In contrast, photonics offers bandwidth in the THz range [1, 2]. However, conventional photonic elements, such as optical fibers, require physical dimensions on the order of the wavelength of light (about a micron) due to diffraction limitations. The difference in physical size between nanometer-scale electronics and micrometer-scale photonic elements yields an incompatibility between the two types of devices. Plasmonics merges the high bandwidth offered by photonics and the nano-scale integration offered by nanoelectronics by coupling a photon's energy with a free-electron gas, creating a subwavelength, oscillating mode known as a plasmon [3-7]. Because plasmonic devices are capable of subwavelength confinement, plasmonics forms the basis of the research area of nanophotonics. Plasmonics and the recent birth of metamaterials [8, 9] (for recent review on optical metamaterials see [10]) and Transformation Optics (TO) [11-14] are currently driving the development of a family of novel devices with unprecedented functionalities such as subwavelength waveguides [5, 15, 16], optical nanoantennas [17-26], superlenses [9, 27-30], optical invisibility cloaks [12, 31-34], hyperlenses [35-38], planar magnifying hyperlens and light concentrators [13, 39, 40].

Because the plasmon phenomenon in optical and telecommunication frequencies typically originates from the collective oscillations of free charges in a material due to an applied electromagnetic field, plasmonic devices generally require metallic components, which have an abundance of free electrons. These free electrons provide the negative real permittivity that is an essential property of any plasmonic material. However, metals are plagued by large losses, especially in the visible and ultra-violet (UV) spectral ranges, arising in part from interband electronic transitions. Even the metals with the highest conductivities suffer from large losses at optical frequencies [41, 42]. These losses are detrimental to the performance of plasmonic devices, seriously limiting the feasibility of many plasmonic applications. In an attempt to mitigate material losses, optical gain materials can be combined with metallic structures [43-57]. However, even the best gain materials available are barely enough to compensate the losses in the metal. Because these losses are inherent to the constituent materials, alternative plasmonic materials with lower losses are required to develop robust plasmonic devices. Lower losses in such improved plasmonic components can be readily compensated with existing gain materials.

Plasmonics could have a large impact on applications at telecommunication and optical frequencies, and hence we begin this review with a survey of several potential plasmonic materials. We compare the relative merits of these potential materials within the context of novel plasmonic devices to gain insight into the suitability of each material in particular applications. In Section 2, we provide a brief background on the various electromagnetic losses associated with solids at frequencies in and near the visible range. In Section 3, we review the optical properties of various plasmonic materials and also discuss the methodologies adopted in choosing potential low-loss materials for plasmonics. Specifically, sections 3.1-3.4 review metals, metal-alloys, heavily doped wide-band semiconductors, and graphene respectively. We formulate the figures-of-merit (or quality factors) for various classes of plasmonic devices to effectively compare the performances of plasmonic materials in Section 4. Finally, in Section 5, we present a comparative study of the performance of the various potential materials based on figures-of-merit and other practical criteria. We conclude the paper with a discussion that quantitatively identifies the best of the reviewed material choices for each class of devices for various regions of the visible and near-infrared (NIR) ranges.

## 2. Background

Polarization describes a material's interaction with electromagnetic waves. While polarization can be electrical and/or magnetic in nature, the magnetic polarization of naturally occurring materials is negligible for frequencies higher than several hundred THz. The electrical polarization can be described by the material's complex electrical permittivity or dielectric function, denoted by $\varepsilon(\omega)$. While the real part of the dielectric function (denoted by $\varepsilon_1$ or $\varepsilon'$) describes the strength of the polarization induced by an external electric field, the imaginary part (denoted by $\varepsilon_2$ or $\varepsilon''$) describes the losses encountered in polarizing the material. Thus, a low loss material is associated with small values of $\varepsilon''$.

Primary loss mechanisms in the NIR, visible, and soft-UV frequencies may be broadly classified as arising from phenomena related to conduction electrons and bound electrons (interband effects) [58]. Losses for conduction electrons arise from electron-electron and electron-phonon interactions, and from scattering due to lattice defects or grain boundaries. Because the conduction electrons have a nearly continuum of available states, their interaction with an electromagnetic field is well approximated by classical theory. The Drude theory [59] describes this phenomenon by treating conduction electrons as a three-dimensional free-electron gas. According to the generalized Drude theory, the permittivity of a material can be written as follows:

$$\varepsilon(\omega)' + i\varepsilon(\omega)'' = \varepsilon(\omega) = \varepsilon_{\text{int}} - \frac{\omega_{\text{p}}^2}{\omega(\omega + i\Gamma)} \quad , \tag{1a}$$

$$\omega_{\text{p}}^2 = \frac{ne^2}{\varepsilon_0 m^*} \quad . \tag{1b}$$

In Eq. (1a), $\Gamma = 1/\tau$ where $\tau$ is the mean relaxation time of conduction electrons, and $\varepsilon_{\text{int}}$ is a contribution due to interband transitions; it is unity for the case of a perfectly free-electron-gas. The plasma frequency ($\omega_{\text{p}}$) is given by Eq. (1b), where $n$ is the conduction electron density, and the effective optical mass of conduction electrons is $m^*$. In general, $\varepsilon_{\text{int}}$ depends on wavelength (which is typically accounted by including the Lorentz oscillators terms [58]), but for some spectral ranges it can be roughly approximated as constant (see Table 1). Also, $\Gamma$ can depend on the size of the plasmonic particle. According to the classical theory, the total damping rate, Γ, is the sum of damping rates due to electron-electron scattering, electron-phonon scattering and lattice defects or grain-boundary scattering. The boundary scattering rate depends on the size of the plasmonic particle, so that the relaxation rate for a spherical particle of size R can be approximated as:

$$\Gamma = \Gamma_{\infty} + A\frac{v_{\text{F}}}{R} \quad , \tag{2}$$

where $\Gamma_{\infty}$ is the relaxation constant of the bulk material, $v_{\text{F}}$ is the Fermi velocity; $A$ depends on details of the scattering process, and is typically on the order of unity [58, 60, 61]. For simplicity, below we assume that $\Gamma = \Gamma_{\infty}$.

**Table 1** Drude model parameters for metals. $\omega_{\text{int}}$ is the frequency of onset for interband transitions. Drude parameters tabulated are not valid beyond this frequency.

| | $\varepsilon_{\text{int}}$ | $\omega_{\text{p}}$ (eV) | $\Gamma$ (eV) | $\omega_{\text{int}}$ (eV) |
|---|---|---|---|---|
| Silver [41, 62, 63] | 3.7 | 9.2 | 0.02 | 3.9 |
| Gold [41, 63] | 6.9 | 8.9 | 0.07 | 2.3 |
| Copper [41, 62, 63] | 6.7 | 8.7 | 0.07 | 2.1 |
| Aluminum [64, 65] | 0.7 | 12.7 | 0.13 | 1.41 |

Because plasmonic applications require materials with negative $\varepsilon'$, Eq. (1a) clearly indicates that this requirement is satisfied for materials with a plasma frequency higher than the desired frequency of application. Because metals tend to have large plasma frequencies and high electrical conductivity, they have traditionally been the materials of choice for plasmonics. In Table 1, we summarize the material parameters for high-conductivity metals as reported in the literature. Among the metallic elements, silver has the smallest $\Gamma$ and is the best-performing choice at optical frequencies. Gold, which has a larger $\Gamma$ than silver, is often the metal of choice at lower NIR frequencies, having the advantage of being chemically stable in many environments. However, gold has high interband losses in the visible spectrum for wavelengths below or about 500 nm. Similarly, copper is plagued by large interband losses over most of the visible spectrum. Thus, silver and gold have predominately been the materials of choice for plasmonic applications around the optical frequencies. However, future plasmonic applications demand even lower losses to fully exploit their potential.

Interband transitions, which form a significant loss mechanism in materials at optical frequencies, occur when electrons jump to higher, empty energy levels caused by absorption of incident photons. In metals, when a bound electron absorbs an incident photon, the electron can shift from a lower energy level to the Fermi surface or from near the Fermi surface to the next higher empty energy level. Both of these processes result in high loss at optical frequencies.

In semiconductors and insulators, valence electrons absorbing the energy from a photon shift into the conduction band, resulting in loss. This loss manifests as an increase in $\varepsilon''$, and can be treated using the formalism of the Lorentz oscillator model. The two-level description of the absorption process results in a simple Lorentz model given by Eq. (3) [66]:

$$\varepsilon_{\mathrm{lk}}(\omega) = \frac{f_{\mathrm{lk}}\,\omega_{\mathrm{p,lk}}^2}{\omega_{\mathrm{lk}}^2 - \omega^2 - i\omega\Gamma_{\mathrm{lk}}} \quad . \tag{3}$$

Here, $f_{\mathrm{lk}}$ corresponds to the strength of the oscillator at energy levels $l$ and $k$, $\omega_{\mathrm{lk}}$ is the resonant frequency corresponding to the difference between the energies of levels $l$ and $k$, $\Gamma_{\mathrm{lk}}$ is the damping in the oscillator accounting for non-zero line-width of the peak, and $\omega_{\mathrm{p,lk}}$ is similar to the plasma frequency given by Eq. (1b) with the difference that $n$ here refers to the concentration of electrons in the lower occupied state. When there are many of such interacting energy levels, the effective permittivity can be expressed as a summation over all allowed Lorentzian terms. This is a popular approach utilized in the Drude-Lorentz model to reasonably approximate the dielectric function of metals [67]. In general, solids with a periodic lattice have electronic energy levels which exist as bands instead of discrete levels, requiring a Joint-Density-of-States description and integration over all the allowed transitions at any given photon energy (a more detailed discussion of this formulation is found in reference [66]).

To illustrate the impact of interband transitions on losses in a material, Fig. 1 shows the imaginary part of permittivity for copper. The free-electron and interband transition contributions to loss are shown along with the values extracted from experimental results [41]. The peak in the experimental $\varepsilon''$ curve at 2.1 eV clearly corresponds to interband transitions; this peak is the result of electronic transitions from the filled $L_3$ band to the Fermi surface [62]. Similarly, the peak at 5.2 eV is due to $X_5 \rightarrow X_{4'}$ transitions and $L_2 \rightarrow L_1$ transitions [62]. Table 1 shows the frequency ($\omega_{\text{int}}$) that marks the onset of significant loss due to interband transitions in several selected metals.

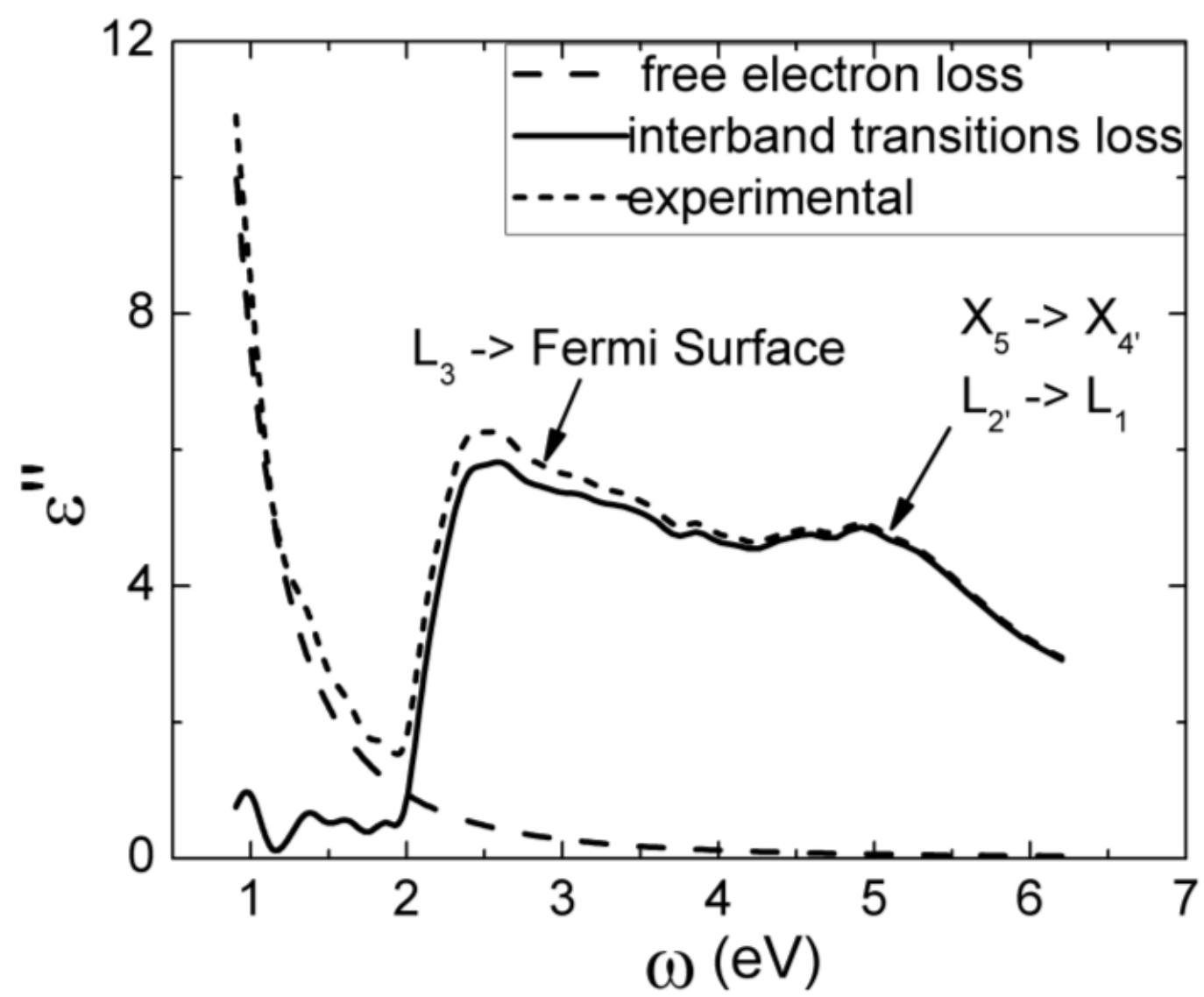

**Figure 1** The losses in Cu shown as the sum of interband losses and free electron losses [41,62]. Annotations identify the interband transitions responsible for peaks in $\varepsilon''$.

The direct transitions discussed above form an important loss mechanism in both semiconductors and insulators. In direct bandgap semiconductors, the bandgap corresponds to the onset of interband transitions. In indirect bandgap semiconductors, indirect transitions contribute to loss for photon energies exceeding the bandgap before the direct transitions set in. For photon energies below the bandgap, there can be appreciable losses due to various mechanisms such as trap-assisted transitions (*e.g.* nitrogen levels in GaP:N [68]), generation of excitons (especially for wide bandgap semiconductors and insulators [69-71]) and transitions between impurity levels [72].

**3. Survey of alternative plasmonic materials**

This section focuses on potential candidates for low-loss plasmonic materials in the NIR, visible, and near-UV spectral ranges. Various metals, metal-alloys, metallic compounds and semiconductors that can offer advantages as plasmonic materials are identified and discussed in terms of their relevant properties. Discussion is categorized into subsections based on the class or type of material reviewed. Metals are considered first, followed by metal-alloys in the second subsection. The third sub-section focuses on doped semiconductors, while the last discusses graphene.

*3.1. Metals as candidates for plasmonics*

As discussed in Section 2, metals are candidates for plasmonic applications because of their high conductivity. Among metals, silver and gold are the two most often used for plasmonic applications due to their relatively low loss in the visible and NIR ranges. In fact, almost all of the significant experimental work on plasmonics has used either silver or gold as the plasmonic material. Silver has been used for the demonstration of a superlens [28, 29], a hyperlens [37], a negative-refractive-index material in the visible range [73], and extraordinary optical transmission [74]. Gold was used for the first demonstration of a negative-refractive-index material in the NIR [75], many studies on Surface-Enhanced Raman Scattering (SERS), the fabrication of plasmon waveguides, and numerous Localized Surface Plasmon Resonance (LSPR) sensors (see for instance, [6]). While metals other than silver and gold have been used in plasmonics, their use is quite limited, as their losses are higher than those of silver and gold. For instance, platinum and palladium have been used as plasmonic materials in systems where the catalytic activity of the plasmonic material is important to the overall device functionality [76, 77]. In addition, nickel films have been reported to have surface plasmon-coupled chemiluminescence, which may merit the use of nickel in particular plasmonic applications.

Among the alkali metals, sodium and potassium have the lowest losses [78]. In fact, these losses are comparable or even better than that of silver. Although there have been several theoretical studies on alkali metals [79], they will not be discussed in this section, as their potential has not been experimentally verified to our knowledge. In pure elemental form, these alkali metals are very reactive to air and water, and therefore they must be stored in mineral oil or Ultra High Vacuum (UHV) environments to avoid highly energetic and dangerous reactions. Such extreme restrictions have made fabrication with alkali metals prohibitive. While depositing alkali metals can be a straightforward process, accomplished with an alkali metal dispenser, other fabrication and characterization obstacles must be overcome before alkali metals find use in plasmonics. In the following section, we will discuss low-loss noble metals (silver, gold and copper) and aluminum.

The plots of $\varepsilon'$ and $\varepsilon''$ of these four metals along with potassium and sodium are shown in Fig. 2.

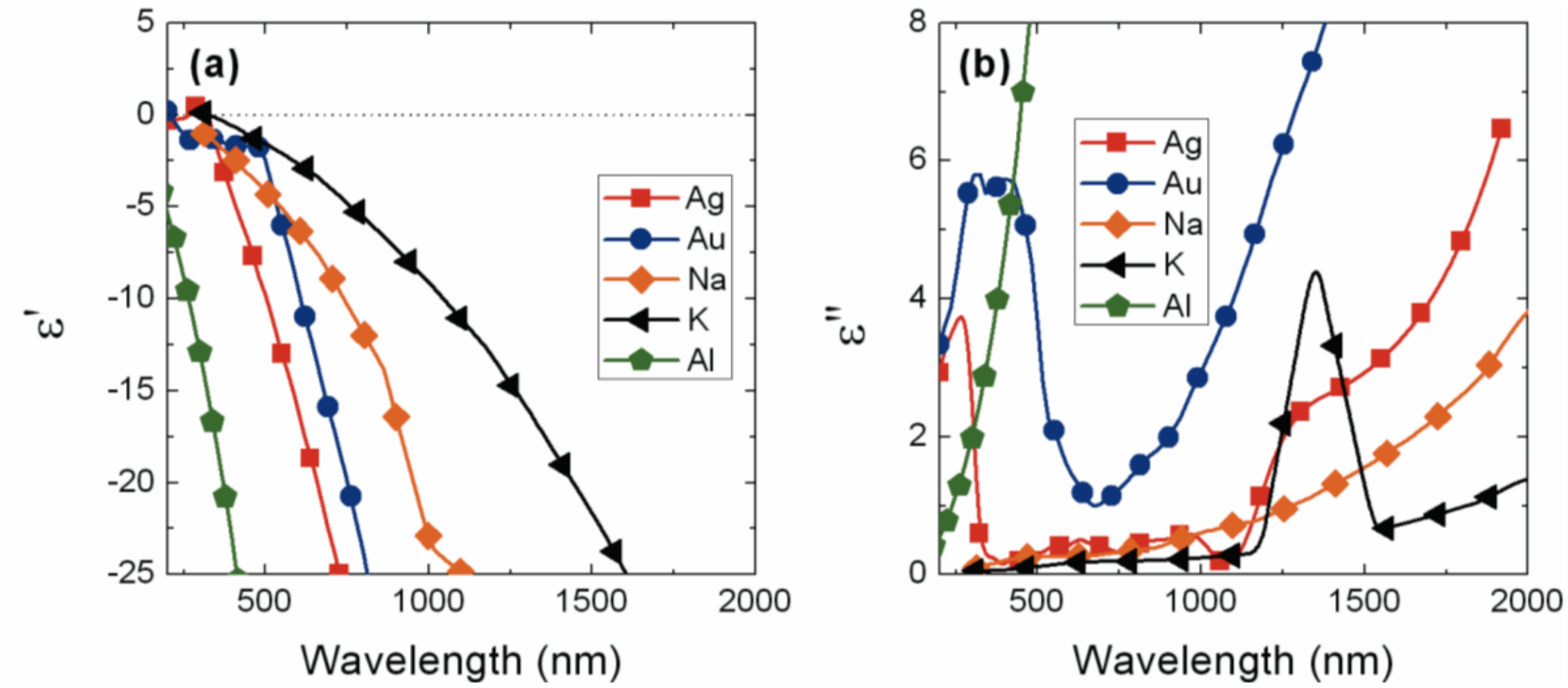

**Figure 2** (color online) Real (a) and imaginary (b) parts of the permittivities of Ag, Au, Na, K, and Al. The data for Ag, and Au are taken from [41], and those of Al, Na, and K are from [78].

As previously mentioned, silver has the lowest loss in the visible and NIR ranges. However, in terms of fabrication, silver degrades relatively quickly and the thickness threshold for uniform continuous films is around 12-23 nm [80-82], making silver less suitable for TO devices (discussed in Section 5). Additionally, silver losses are strongly dependent on the surface roughness [58]. Gold is the next-best material in terms of loss in the visible and NIR ranges. Compared with silver, gold is chemically stable and can form a continuous film even at thicknesses around 1.5-7 nm [83-85]. It is very important to note that films deposited with thicknesses below the percolation threshold are semi/discontinuous with drastically different optical properties compared to uniform films [86, 87]. These fabrication complications make these materials undesirable at thicknesses below the percolation threshold when continuous films are needed. Silver and gold films can be fabricated by various physical vapor deposition (PVD) techniques and nanoparticles and metal-coated nanoparticles can be synthesized by liquid chemical methods. Typical PVD methods include electron-beam/thermal evaporation and sputtering. In liquid chemical methods, chloroauric acid ($H[AuCl_4]$) and silver oxide ($Ag_2O$)/silver nitrite ($AgNO_4$) are commonly used for gold and silver, respectively.

Because copper has the second-best conductivity among metals (next to silver), it is expected to exhibit promising plasmonic properties. Indeed, $\varepsilon''$ of Cu is comparable to that of Au from 600-750 nm. Considering the cost of silver and gold, copper would be a good candidate to replace silver and gold as a plasmonic material if the performance of copper were tolerable. Unfortunately, fabricating devices with copper is challenging, as it easily oxidizes and forms $Cu_2O$ and CuO. A systematic study of the oxidation effects on copper for LSPR modes is found in [88]. In their results, Chan *et al.* have demonstrated that oxide-free nanospheres exhibit a sharp and narrow LSPR peak comparable to that of silver and gold.

Aluminum has not been an attractive plasmonic material due to the existence of an interband transition around 800 nm (1.5 eV), resulting in large $\varepsilon''$ values in the visible wavelength range (see Fig. 2). However, in the UV range, $\varepsilon'$ is negative even at wavelengths smaller than 200 nm where $\varepsilon''$ is still relatively low. Thus aluminum is a better plasmonic material than either gold or silver in the blue and UV range. It is important to note that the $\varepsilon'$ values of silver and gold do not become positive until wavelengths greater than 326 nm and 207 nm, respectively. Aluminum is easily oxidized and very rapidly forms an aluminum oxide ($Al_2O_3$) layer under atmospheric conditions, making device fabrication with aluminum challenging. The thickness of this $Al_2O_3$ layer is typically 2.5-3 nm [89], and the presence of this oxide layer results in a red shift in LSPR peak position [90]. Despite these challenges, aluminum has been used in plasmonic systems in the UV-blue spectral region such as to study LSPR [89, 90], surface plasmon polariton (SPP) propagation [91], surface-enhanced fluorescence [92, 93], and Raman spectroscopy [94, 95].

### *3.2. Metallic Alloys*

Metallic alloys, intermetallics and metallic compounds are potential candidates for alternative plasmonic materials owing to their large free electron densities. This section primarily discusses the various techniques employed to tune the optical characteristics of metals by making alloys. Metallic compounds such as magnesium diboride ($MgB_2$) are not discussed here due to their poor performance as plasmonic materials around optical frequencies [96-100]. Similarly, intermetallics such as silicides, which are reported to be plasmonic materials, will not be discussed owing to their large losses in the NIR and visible spectrum [101-107].

#### 3.2.1. Noble-Transition Alloys

Because of the strong plasmonic performance of noble metals, one approach to improve these materials is by shifting their interband transitions to another (unimportant) part of the spectrum. This can be achieved by alloying two or more elements to create unique band structures that can be fine-tuned by adjusting the proportion of each alloyed material.

Noble-transition metal alloys are one interesting set of potential alternative plasmonic candidates. Bivalent transition metals, such as Cadmium and Zinc, when doped into monovalent noble metals contribute one extra electron to the free-electron plasma [108]. This results in n-type doping of noble metals, which raises the Fermi level, increases $\omega_p$, and shift the threshold for interband transitions, thereby modifying the optical spectra of the alloy. By changing the Fermi level of a metallic layer, one can, in principle, reduce the absorption at a specific wavelength. This could be accomplished, for example, by shifting the Lorentzian peak to some other wavelength that is unimportant for a particular application. This process of manipulating and fine-tuning a material's electronic band structure to achieve desirable electronic properties is referred to as "band engineering."

The particular experiment detailed in [108] involved n-doping gold with cadmium and employed samples with pre-determined stoichiometries that were purchased from commercial vendors. Alloying gold and cadmium creates a unique band structure, shifting the peak losses to new frequencies and resulting in higher losses in one frequency range, with simultaneously lower losses in another range. To illustrate this point, Fig. 3 shows a simulation demonstrating how doping gold with cadmium shifts the interband transition peaks, confining high losses to one region while lowering losses almost everywhere else in the spectrum. This method of raising the Fermi level by small amounts can continue for low doping levels (<10% volume), but the technique will break down as the doping levels become high enough to significantly modify the material's band structure. This technique may be extended to different material combinations, creating unique band structures optimized for specific frequencies and applications.

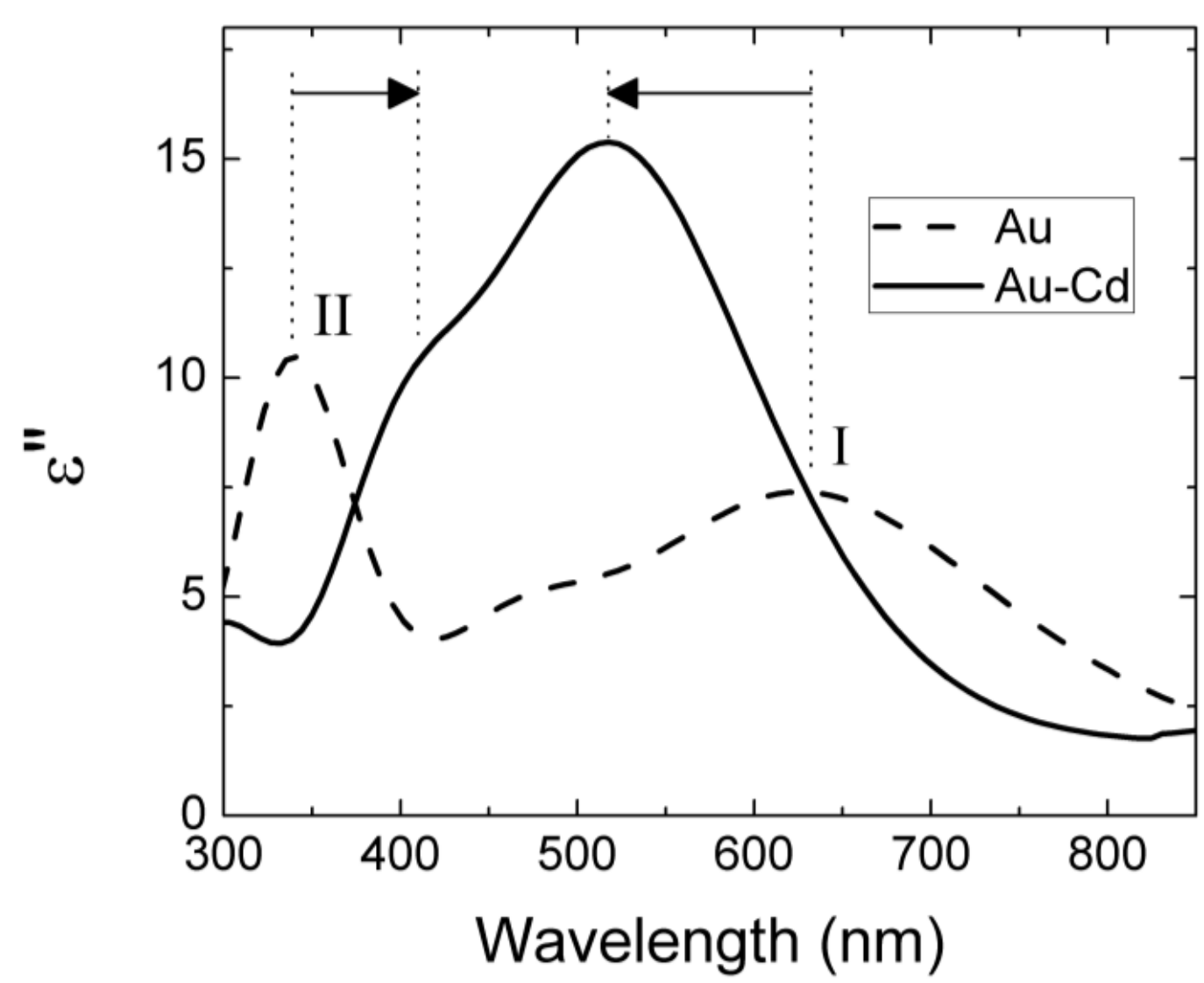

**Figure 3** (By courtesy of Dr. M. Noginov and Dr. V. Gavrilenko, Norfolk University [108]) Numerical simulation detailing the shift in interband transition peak and modified

band structure when Au [41] is doped with 3.3 at.% Cd. The dashed curve shows the imaginary permittivity for the undoped gold, while the solid curve is the result for the doped alloy. The doping process combines the losses from peaks I&II into a single, confined high-loss region, while leaving lower losses elsewhere.

3.2.2. Alkali-Noble Inter-Metallic Compounds

Alkali-noble metal inter-metallic compounds are another group of candidates for low-loss metals because the Group I alkali metals exhibit the strongest free-electron-like behavior [109]. As far back as 1978 [110], the permittivity of $Li_2AgIn$ was experimentally measured, exhibiting zero-loss permittivity values at 2 eV without mention of the implication of this result for plasmonic devices. Note that this paper presents data showing negative losses, which raises some doubts about the accuracy of measurements or parameter extraction. While many alkali-noble metal binary compounds have been presented as alternative plasmonic materials [109], the compound predicted to show the most promise is potassium gold (KAu), which has been calculated to have zero interband losses below its unscreened plasma frequency at 1.54 eV. However, because of Drude losses, the crossover point is 0.5 eV. Because KAu does not have negative $\varepsilon'$ values in the telecommunication and visible frequency ranges, it cannot be considered a plasmonic candidate for applications in these spectral ranges.

Fabrication of alkali-noble compounds can be challenging. Just bringing two metals of significantly different properties together to form an alloy can be a non-trivial task. The vapor pressures of potassium and gold differ by 10 orders of magnitude and their surface energies differ by an order of magnitude, making the alloying process difficult. Assuming these compounds behave similarly to pure alkali metals, these highly reactive alloys must either be characterized in-situ or be passivated prior to removing them from the fabrication chamber in order to prevent rapid oxidation. Passivation would allow for further optical characterization of the fabricated sample using ellipsometry, prism coupling, SPP propagation, and other techniques, but the passivation itself may alter the surface properties of the sample as well as limit applicability of such films for real device fabrication. Because such compounds have not been extensively studied in the past, the phase, stoichiometry and growth kinetics of these alloy systems are not well understood.

*3.3. Semiconductors*

Semiconductors are conventionally regarded as dielectric materials for frequencies above several hundred THz. However, semiconductors can actually exhibit a negative real permittivity in this spectral region under certain circumstances [111-114]. Due to the ease of fabrication and flexibility in tuning their properties such as carrier concentration, semiconductors are also potential materials for plasmonics. In order to qualify as a low-loss plasmonic material, the bandgap and plasma frequency of the semiconductor both must be larger than the frequency range of interest. While a large plasma frequency ensures a negative real permittivity, a large bandgap ensures almost no interband transition losses. Semiconductors can exhibit negative $\varepsilon'$ in IR frequencies when heavily doped [111-113] or in resonance (*e.g.* phonon resonance in silicon carbide [78, 114, 115]). Because resonance is a narrow-band phenomenon and phonon resonance occurs at low frequencies, here we consider only heavily doped semiconductors as possible low-loss plasmonic materials. Thus, a wide bandgap, heavily doped semiconductor with high carrier mobility can qualify as a low loss plasmonic material around the optical frequencies.

Despite the abundance of semiconductors with large bandgap values (> 1.5 eV) and high carrier mobilities, very high doping levels are necessary to bring the crossover frequency of semiconductors into the optical range, and achieving these doping levels is challenging. Hoffman *et al.* [113] reported that doping gallium arsenide to $7 \times 10^{18}$ $cm^{-3}$ can raise the crossover frequency (where $\varepsilon'$ changes from negative to positive) to about 9 µm. However, to bring the crossover frequency near the optical range, a doping level of at least $3 \times 10^{20}$ $cm^{-3}$ is required. The necessity of doping semiconductors so heavily raises concerns about the solid solubility limit, the fraction of dopants that would be active, and doping compensation effects [116, 117]. Another major concern at such high doping levels is retaining the high carrier mobility that is essential for low losses. Due to these issues, plasmonics in the optical spectrum has remained mostly out-of-reach for semiconductor materials.

However, indium-tin-oxide (ITO) has been shown to be a potential plasmonic material in the NIR region [118-123]. ITO is a transparent, conducting oxide typically consisting of 90%wt indium oxide ($In_2O_3$) and 10%wt tin oxide ($SnO_2$). ITO has been widely studied in the field of optoelectronics. Because ITO is non-stoichiometric, predictions show its plasma frequency can be engineered between 0.44 eV and 6.99 eV by varying the tin (Sn) doping level in $In_2O_3$ up to 45%wt [124]. Experimentally, the plasma frequency has been measured between 0.78 and 2.13 eV [112, 125, 126]. Robusto *et al*. have demonstrated SPPs in ITO between 1.8 µm and 1.9 µm [118], which was followed by many other groups [119-123] reporting SPPs in the NIR region. Therefore, ITO appears to be an appealing plasmonic material in the NIR and optical frequency ranges.

Because ITO is a non-stoichiometric compound, its optical properties largely depend on the growth/deposition processes and annealing conditions, including the temperature and ambient gasses [112]. Sputtering and laser ablation techniques have reliably produced quality films of ITO [127]. Post-deposition annealing in nitrogen has been shown to produce more conductive and less transparent films due to increased oxygen vacancy ($V_o$) defects [128]. On the other hand, less conductive and more transparent films are produced by annealing ITO in an oxygen ambient [128]. In our own experiments, we have observed significant changes in the dielectric function of e-beam evaporated ITO films annealed in different ambient environments, as shown in Fig. 4. While ITO films annealed in oxygen do not show negative $\varepsilon'$ in the spectrum of interest, the films annealed in nitrogen ambient do show negative real permittivity. Figure 4 also shows that ITO's properties depend on the annealing temperature. A higher annealing temperature increases the conductivity of the ITO film and reduces the loss within the wavelength range of interest. It is important to note that the loss in ITO films is comparable to that of silver in the NIR range but is significantly lower than the losses in silver at longer wavelengths.

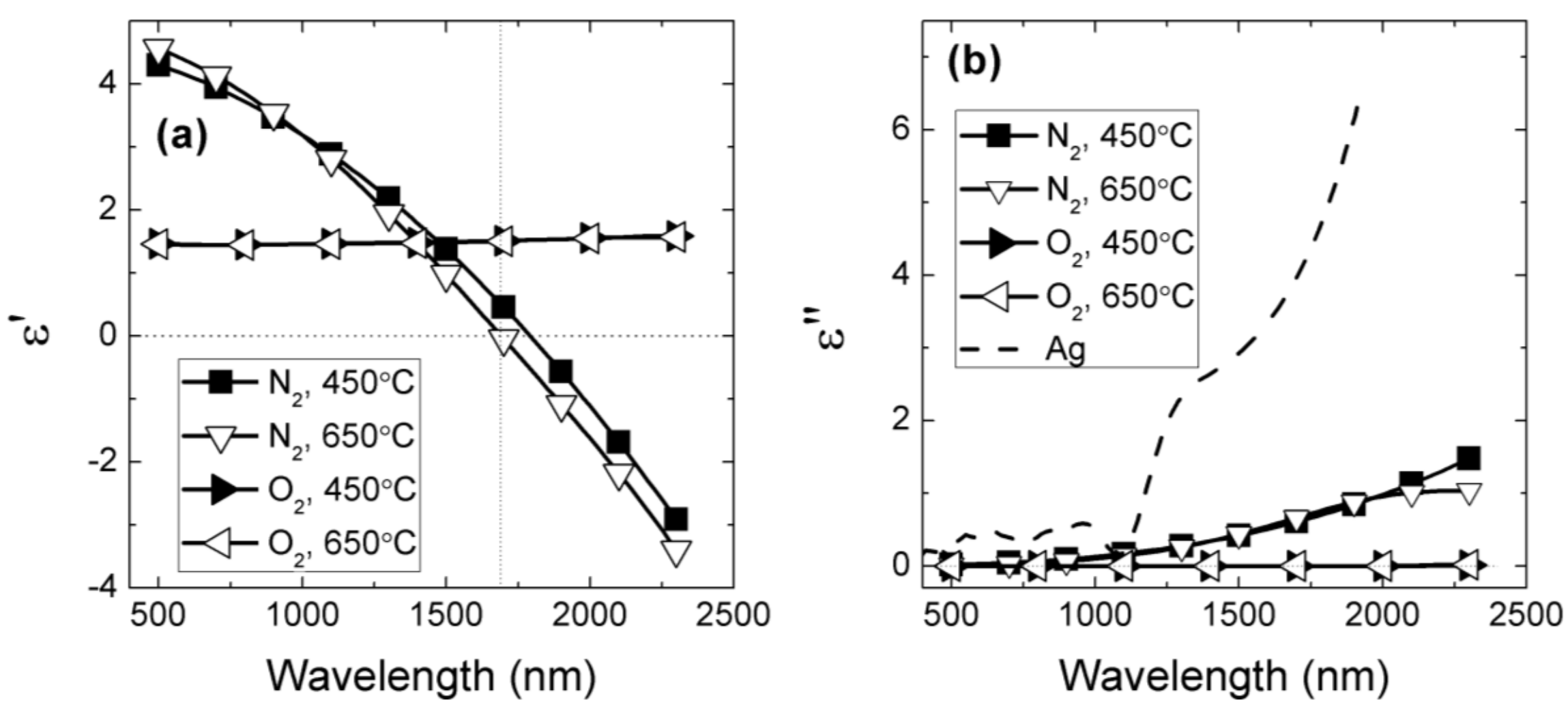

**Figure 4** Real (a) and imaginary (b) parts of permittivity of ITO annealed at various conditions (N2, 450 °C; N2, 650 °C; O2, 450 °C; O2, 650 °C).

Other than ITO, high performance conductive zinc oxides such as aluminum-zinc-oxide (AZO) and gallium-zinc-oxide (GZO) can be promising low-loss alternatives in NIR. These conductive zinc oxides are widely studied for applications in transparent electronics. Based on the optical characteristics of these films reported in literature, we have found that AZO and GZO can have significantly lower loss than silver at telecommunication wavelengths which are of particular importance for photonics and nanophotonics applications (see Fig. 5) [129-131]. While AZO can exhibit losses more than three times lower than that of silver at the wavelength of 1.5 μm, GZO being only slightly inferior to AZO, can exhibit lower losses too.

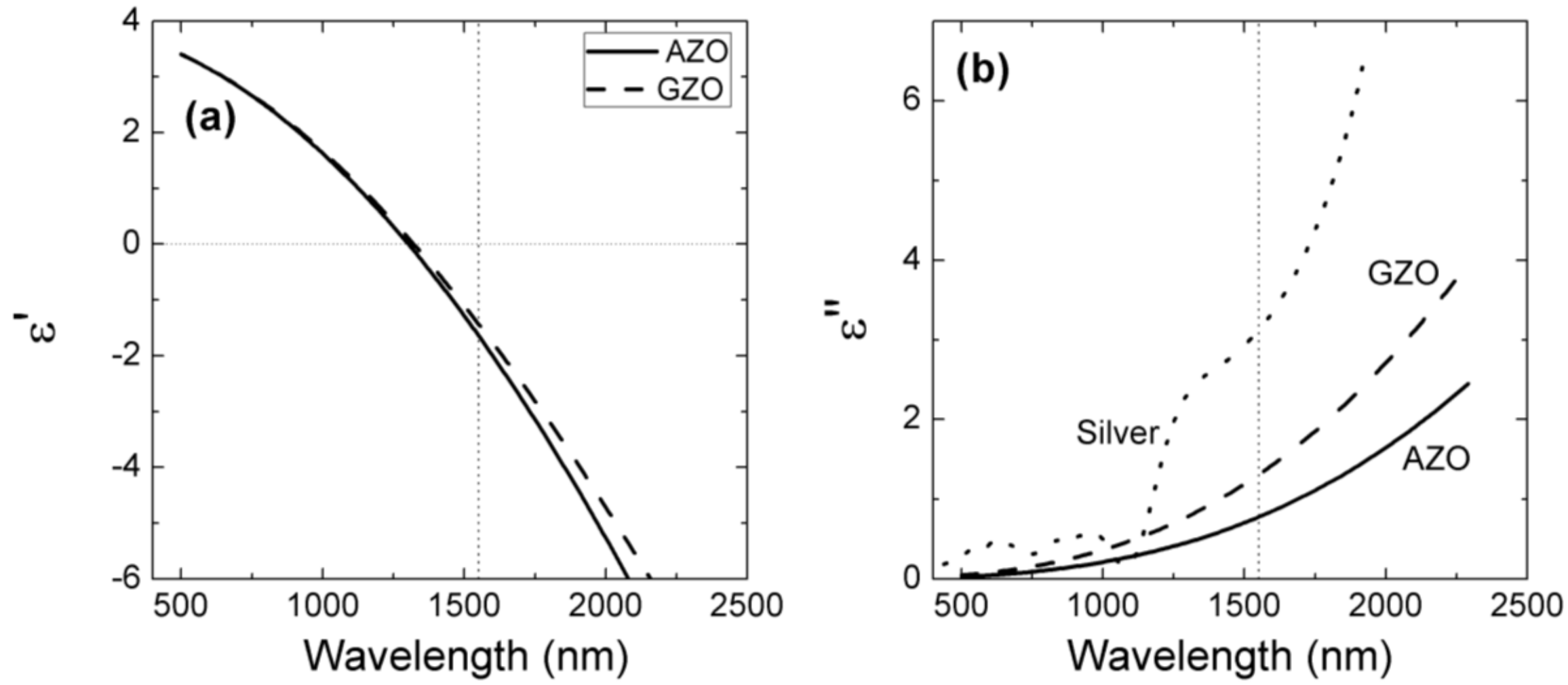

**Figure 5** Real (a) and imaginary (b) parts of permittivity of AZO and GZO obtained from parameters reported in reference [129]. The losses in AZO and GZO are much smaller than that of silver [41] at the telecommunication wavelength.

Similar to other conductive oxides, AZO and GZO are non-stoichiometric, and deposition conditions play a crucial role in achieving the desired properties. High quality thin films of AZO and GZO may be produced by sputtering or laser ablation. Deposition temperature, oxygen partial pressure during deposition and dopant concentration must be optimized in order to achieve low loss as well as negative real permittivity [111, 129-133]. For example, M. H. Yoon *et al.* report that large doping of zinc oxide does not necessarily increase the carrier concentration [134]. In AZO, a large doping level results in the formation of $ZnAl_2O_4$, which does not increase carrier concentration, but adversely affects the carrier mobility through increased scattering. This causes the zero-cross-over of real permittivity to shift towards longer wavelengths. We observe this happening in Pulsed Laser Deposited (PLD) AZO films with 2%wt $Al_2O_3$. Evidently, lower $Al_2O_3$ concentration of about 0.8%wt produces AZO films with $\varepsilon'$-cross-over at wavelength smaller than 1.5 μm [129]. With losses being much smaller than in silver at this wavelength, AZO is a promising low-loss alternative material for plasmonics in NIR.

Aside from oxide semiconductors, III-V semiconductors (*e.g.* GaAs, GaN, GaP) and SiC are potential candidates for plasmonics in the NIR and optical spectral ranges. However, heavy doping could again be a problem that must be addressed when considering these materials as options for alternative plasmonic materials.

*3.4 Graphene*

Graphene is another material that has generated excitement in the research community due to its unique band structure and high carrier mobility [135-138]. Graphene is a two-dimensional system - enabling excitation of surface plasmons (SP) similar to the surface plasmons on metal/dielectric interfaces. However, these two-dimensional plasmons in graphene exhibit a dispersion relation different compared to plasmons in three dimensions [139-141]. Some schemes for plasmon based THz oscillators in graphene have already been proposed in literature [142-144]. These plasmon modes can possibly exist up to near-IR frequencies. However, in order to evaluate the potential of graphene as an alternative low loss plasmonic material, the losses in the plasmon modes must be evaluated [145].

It can be expected that interband transitions contribute significantly to losses because of the zero band-gap in graphene. These interband transitions occur above a threshold determined by Fermi energy and plasmon wavevector [140]. However, this threshold can be shifted to frequencies beyond the region of interest by larger doping, which increases the Fermi energy. It has already been experimentally demonstrated that the interband threshold can be tuned by varying electrical doping [146]. Below the interband threshold, the losses are primarily due to impurity scattering and excitation of optical phonons together with electron-hole pairs. Jablan *et al.* have analyzed the electron relaxation times due to different loss mechanisms in graphene, and demonstrate that graphene may inherently contain lower losses relative to conventional metal/dielectric interfaces up to frequencies corresponding to 0.2 eV [145]. Initial theoretical estimates indicate that graphene is a good plasmonic material for THz applications. However, at NIR frequencies, losses in graphene may still be comparable to noble metals. This makes graphene less attractive as an alternative plasmonic material at the telecommunications and visible wavelengths.

**4. Quality Factors**

Quality factors, or figures-of-merit, form a common platform to compare the performances of various materials used in different applications over a wide frequency band. Although the loss in a material characterized by $\varepsilon''$ is a necessary indicator of performance, the real part of permittivity $\varepsilon'$ is also important in quantifying the overall material quality in many devices. Because the field distribution in a material depends on $\varepsilon'$ and the loss depends on $\varepsilon''$, performance metrics or quality factors for a plasmonic material are generally a function of both $\varepsilon'$ and $\varepsilon''$. Considering that different applications can have different definitions for the quality factors [147-151], our discussion is focused on four major classes of plasmonic devices: LSPR based devices,

SPP waveguides, TO devices and superlens. It is important to note that the following discussions are valid only in the range of frequencies for which $\varepsilon'$ is negative. For metals and metal-like materials such as heavily doped semiconductors, this translates to frequencies below the crossover frequency. It should also be noted that in the following discussion, the term "metal" is used for both metals and metal-like materials for brevity.

In the first part of this section, the quality factors of LSPR and SPP systems are considered and are denoted as $Q_{\mathrm{LSPR}}$ and $Q_{\mathrm{SPP}}$, respectively. LSPR and SPR systems produce local-field enhancement at the surface of metallic components [148]. Hence their quality factors can be defined as follows:

$$Q_{\mathrm{f}} = \frac{\text{(Enhanced local-field)}}{\text{(Incident field)}} \quad . \tag{4}$$

Quality factor for LSPR depends significantly on the shape of the metal nanoparticles. For a sphere, $Q_{\mathrm{LSPR}}$ is given by Eq. (5):

$$Q_{\mathrm{LSPR}}(\omega) = \frac{-\varepsilon'(\omega)}{\varepsilon''(\omega)} . \tag{5}$$

For a cigar-shaped spheroid, Eq. (5) becomes the formula below [148]:

$$Q'_{\mathrm{LSPR}}(\omega) = \frac{\varepsilon'(\omega)^2}{\varepsilon''(\omega)} \quad . \tag{6}$$

For SPR, quality factor $Q_{\mathrm{SPR}}$ assumes the same form as Eq. (5). $Q_{\mathrm{SPP}}$ can be defined as the ratio of the real part of the propagation wavevector ($k'_{\mathrm{x}}$) to the imaginary part ($k''_{\mathrm{x}}$) [152]:

$$Q'_{\mathrm{SPP}}(\omega) = \frac{k'_{\mathrm{x}}(\omega)}{k''_{\mathrm{x}}(\omega)} = \frac{\varepsilon'_{\mathrm{m}}(\omega) + \varepsilon_{\mathrm{d}}(\omega)}{\varepsilon'_{\mathrm{m}}(\omega)\varepsilon_{\mathrm{d}}(\omega)} \frac{\varepsilon'_{\mathrm{m}}(\omega)^2}{\varepsilon''_{\mathrm{m}}(\omega)} , \tag{7}$$

where $\varepsilon_m$ is the permittivity of the metal, and $\varepsilon_{\mathrm{d}}$ is the permittivity of the surrounding dielectric material. If $|\varepsilon_{\mathrm{m}}'| >> \varepsilon_{\mathrm{d}}$, Eq. (7) can be simplified as follows:

$$Q_{\mathrm{SPP}}(\omega) = \frac{\varepsilon'_{\mathrm{m}}(\omega)^2}{\varepsilon''_{\mathrm{m}}(\omega)} . \tag{8}$$

It may be noted that $Q_{\mathrm{SPP}}$ has the same form as $Q'_{\mathrm{LSPR}}$ defined for spheroid nanoparticles. The quality factors for LSPR and SPP are shown in Fig. 6. This figure does not include semiconductors and alloys, as their quality factors are much lower than the rest. Equations (5) and (8) explain how $Q_{\mathrm{LSPR}}$ and $Q_{\mathrm{SPP}}$ become large when a material has a large, negative $\varepsilon'$ and a small $\varepsilon''$. These equations clearly convey why conventional plasmonic materials such as silver and gold, which have a large, negative $\varepsilon'$ and a low $\varepsilon''$, have been the plasmonic materials of choice for most applications.

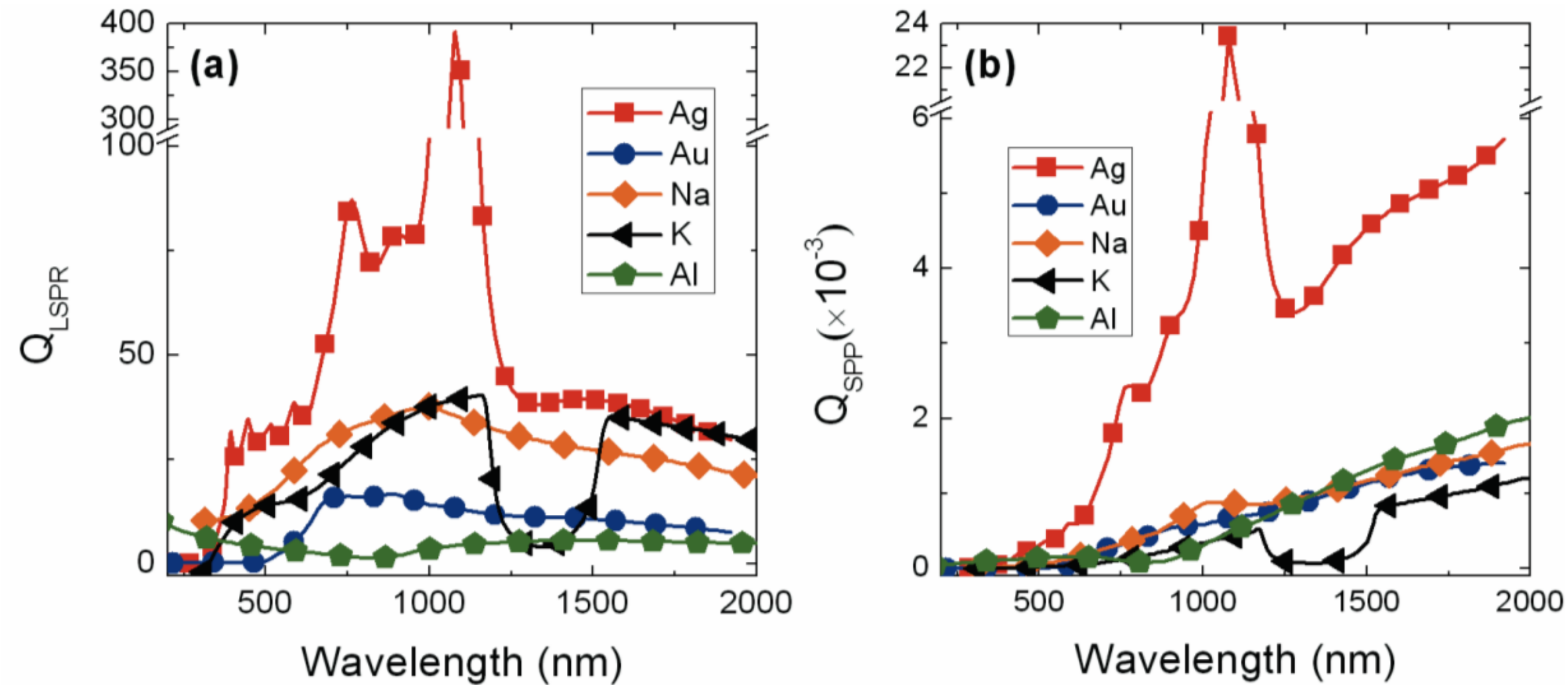

**Figure 6** (color online) Quality factors for localized surface plasmon resonances (QLSPR) is shown in panel (a), and surface plasmon polaritons quality factor (QSPP) is shown in panel (b). The permittivity values used to calculate the presented quality factors are the same as those shown in Fig. 2.

Transformation optics (TO) devices, however, require a different figure of merit due to the nature of TO devices. These devices typically need materials with a real part of *effective* permittivity close to zero at the operating frequency [14]. Thus, TO devices require the response of metallic components to be nearly balanced by that of dielectric components (which typically have $\varepsilon_d \sim 1$). Hence, for practical values of metal filling factors, the magnitude of $\varepsilon'_m$ of metal must be comparable in magnitude (and opposite in sign) to that of the dielectric component. Thus, plasmonic components of TO devices operate near their crossover frequency where $\varepsilon'_m$ is negative and small in magnitude. Thus, in this case, only the losses ($\varepsilon''$) are relevant in defining the quality factor:

$$Q_{TO} = \frac{1}{\varepsilon''_m} \quad (-\varepsilon'_m \sim \varepsilon_d \sim 1) \quad . \tag{9}$$

We now turn our attention to the superlens and its resolution limits. The resolution limit of superlens can be defined through

$$\frac{\Delta}{d} = \frac{-2\pi}{\ln\left(-\frac{\varepsilon''}{2\varepsilon'}\right)} \quad , \tag{10}$$

where $\Delta$ is the minimum resolvable feature size and $d$ is the thickness of the superlens [153]. The value $\Delta/d$ can be considered as the normalized resolution of the superlens. For the purpose of comparative studies using quality factors, we define the inverse of $\Delta/d$ as the superlens quality factor $Q_S$. For the sake of comparison, the host material is assumed to be air, hence $\varepsilon'$ of the superlens is set to -1 for the following discussion. Therefore, $Q_S$ can be expressed as

$$Q_S = \frac{d}{\Delta} = \frac{-\ln\left(\frac{\varepsilon''}{2}\right)}{2\pi} \quad (\varepsilon' = -1) \quad . \tag{11}$$

Quality factors, give a quantitative assessment of the performance of plasmonic materials in their respective categories. However, practical considerations arising from fabrication and integration issues must be considered before choosing the best material.

**5. Comparative studies**

A summary of the figures-of-merit for various materials discussed in the previous sections is provided in Table 2. While quality factors describe how well a material will perform for various applications, there are limiting issues in terms of processing and fabrication. For example, alkali metals are difficult to work with because they are extremely reactive in air ambient environments. Thus, alkali metals have not been used in plasmonic applications regardless of their high quality factors. In addition, silver and aluminum are not ideal materials for the fabrication of plasmonic devices because these materials easily oxidize when exposed to air, which can significantly alter their plasmonic properties. Other fabrication issues can arise in particular designs, such as the formation of extremely thin metallic films and the controlled synthesis of metallic nanoparticles.

**Table 2** Summary of the quality factors of a number of possible low-loss plasmonic candidate materials. The maximum value of the quality factor (up to 2.5 μm) and the wavelength at which the maximum occurs are tabulated for each material and each of the four applications. The quality factors at 1.5 μm (for nanophotonic applications) are also presented. Quality factors for TO devices were calculated at the crossover frequency where $\varepsilon' = 0$.

| | **LSPR & SPR** | | **SPP** | | **TO Devices** | **Superlens** | **Comments** |
|---|---|---|---|---|---|---|---|
| **Material** | **Maximum $Q_{LSPR}$ (λ)** | **$Q_{LSPR}$ (1.5μm)** | **Maximum $Q_{SPP}$ (λ)** | **$Q_{SPP}$ (1.5μm)** | **$Q_{TO}$ (λ)** | **$Q_S$ (λ)** | |
| **Ag*** | 392 (1.08 μm) | 39.3 | 23413 (1.08 μm) | 4530 | 1.82 (326 nm) | 0.3 (339 nm) | Good for LSPR & SPP |
| **Au*** | 16.66 (0.89 μm) | 10.63 | 1410 (1.94 μm) | 1140 | 0.29 (207 nm) | -0.13 (252 nm) | Good for LSPR |
| **Al** | 13.56 (0.113 μm) | 5.55 | 2677 (2.5 μm) | 1315 | 26.32 (82 nm) | 0.52 (114 nm) | Good for LSPR in UV |
| **Na*** | 37.8 (1.00 μm) | 27.3 | 1889 (2.25 μm) | 1179 | NA*** | 0.48 (312 nm) | Difficult to process |
| **K*** | 40.6 (1.17 μm) | 19.2 | 1287 (2.25 μm) | 419 | 22.22 (326 nm) | 0.5 (438 nm) | Difficult to Process |
| **KAu** | 1.3 (2.5 μm) | NA** | 1.1 (2.5 μm) | NA** | 1.72 (2.38 μm) | 0.18 (2.52 μm) | Difficult to process |
| **ITO*** | 2.72 (2.3 μm) | NA** | 16 (2.3 μm) | NA** | 1.54 (1.69 μm) | 0.13 (1.88 μm) | Good for TO in NIR |
| **AZO** | 3.28 (2.26 μm) | 1.46 | 33.1 (2.5 μm) | 2.33 | 2.16 (1.3 μm) | 0.179 (1.46 μm) | Good for TO in NIR |
| **GZO** | 1.8 (2.3 μm) | 0.923 | 15.96 (2.5 μm) | 1.01 | 1.22 (1.32 μm) | 0.087 (1.48 μm) | Good for TO in NIR |

*Ag and Au data ends at 2 μm, Na, K data ends at 2.25 μm, and ITO data ends at 2.3 μm.

**Crossover frequency for these materials corresponds with a wavelength above 1.5 μm. Thus, quality factors are not applicable for these materials at the telecommunication wavelength.

***Crossover frequency data is not available.

Therefore, despite the fact that sodium and potassium have the highest $Q_{LSPR}$ and $Q_{SPP}$ values next to silver, they are extremely difficult to work with, and are therefore probably not practical from a fabrication standpoint. In most cases, silver is by far the best material in terms of quality factor, but it is associated with problems such as oxidation and cost. Aluminum has the advantage of having an extremely high plasma frequency, and it is the only reviewed material that acts as a metal in the UV part of the spectrum. However, aluminum also oxidizes quite easily, which can cause issues in terms of fabrication.

Quality factors for TO devices and superlens indicate that alkali metals can be a good choice. However, they have problems with processing and hence are not the material of choice. Aluminum has a high $Q_{\mathrm{TO}}$ value, but is probably not particularly good for TO applications. This is due to the fact that at the Al crossover wavelength of 81 nm, the thickness of the aluminum layer can be at most 8 nm for effective medium theory to hold. In this range, film roughness, as well as the formation of oxidation layers, could be detrimental to the plasmonic properties of the aluminum structure. However, Aluminum will perform well as a superlens, if these issues can be resolved. Silver also suffers from similar fabrication problems. From the alkali-noble metal alloys, KAu is a possible candidate for TO or superlens applications in terms of the material's quality factor, but there are serious challenges with the synthesis and chemical stability of KAu. In contrast, doped zinc oxide and indium-tin-oxide are realistic choices for TO devices in the NIR range including the telecommunication wavelength, which is particularly important for nanophotonics circuitry. With the optimization of processing conditions, these materials can outperform silver for TO applications, potentially enabling unsurpassed control of light on the nanoscale.

## 6. Conclusions

We have discussed the optical properties of different plasmonic materials, including a comparison of their predicted performance metrics in terms of quality factors. The comparative study shows that there is not a single clear choice for the best low-loss plasmonic material for all applications. Our comparison demonstrates that silver dominates all of the materials we considered in terms of its quality factors in LSPR and SPP applications in the visible and NIR ranges. However, silver is not the clear choice as a low-loss plasmonic material for other applications such as TO and superlens. Even though alkali metals and aluminum have high quality factors for TO devices and superlens, they pose processing challenges. Thus, silver is the best material for a superlens in the near UV. In NIR, AZO may be the best material for TO devices and superlensing, followed by ITO and GZO. Furthermore, these oxide semiconductors can work well at the telecommunication wavelength, which makes them very important substitutes for conventional materials such as gold and silver.

A material's quality factor describes how well a material performs, but it does not give insight into other aspects such as ease of processing and feasibility of integration. A final choice of materials requires a trade-off between quality factor, fabrication practicality, and cost. As an example, the material cost of gold and silver prohibit their wide-scale adoption in cost-driven markets such as photovoltaics.

With the rapid development of nanophotonics, it is clear that there will not be a single plasmonic material that is suitable for all applications at all frequencies. Rather, a variety of material combinations must be fine-tuned and optimized for individual situations or applications. While several approaches and materials have been presented, the problem of losses in plasmonic materials remains open-ended. An improved plasmonic material has the potential to make an enormous impact on both optics and nanoelectronics by allowing for a new generation of unparalleled device applications. We expect the comparative study presented herein to be useful in the elimination of poor choices, and it will serve as a guide in making the optimum choice for a low-loss plasmonic material in various applications.

## 7. Acknowledgments

We would like to thank Martin Blaber and Michael Ford for their interesting discussion on alternative plasmonic materials including alkali-noble alloys along with quality factor derivations. We would also like to thank Joerg Appenzeller, Marin Soljacic, Evgenii Narimanov, and Zubin Jacob for their insight into graphene as an alternative plasmonic material, along with Mikhail Noginov for the data provided in Fig. 3. The authors are also grateful to Professor Sir John Pendry for useful discussions. This work was supported in part by ARO grant W911NF-09-1-0516 and by ARO-MURI award W911NF-09-1-0539.

## 8. References

[1] E. Ozbay, Science **311**, 189-193 (2006).

[2] M. J. Kobrinsky, B. A. Block, J. F. Zheng, B. C. Barnett, E. Mohammed, M. Reshotko, F. Robertson, S. List, I. Young, and K. Cadien, Intel Technol. J. **8**, 129-141 (2004).

[3] R. H. Ritchie, Phys. Rev. **106**, 874-881 (1957).

[4] W. L. Barnes, A. Dereux, and T. W. Ebbesen, Nature **424**, 824-830 (2003).

[5] S. A. Maier and H. A. Atwater, J. Appl. Phys. **98**, 011101 (2005).

[6] S. A. Maier (1st ed), Plasmonics Fundamentals and Applications (Springer, New York, 2007), p. 21-34, 65-88.

[7] S. Lal, S. Link, and N. Halas, Nat. Photonics **1**, 641-648 (2007).

[8] V. G. Veselago, Physics-Uspekhi **10**, 509-514 (1968).

[9] J. B. Pendry, Phys. Rev. Lett. **85**, 3966-3969 (2000).

[10] W. Cai and V. M. Shalaev (1st ed), Optical Metamaterials: Fundamentals and Applications (Springer, New York, 2009).

[11] A. J. Ward and J. B. Pendry, J. Mod. Opt. **43**, 773-793 (1996).

[12] J. B. Pendry, D. Schurig, and D. R. Smith, Science **312**, 1780-1782 (2006).

[13] A. V. Kildishev and V. M. Shalaev, Opt. Lett. **33**, 43-45 (2008).

[14] V. M. Shalaev, Science **322**, 384-386 (2008).

[15] S. A. Maier, P. G. Kik, H. A. Atwater, S. Meltzer, E. Harel, B. E. Koel, and A. A. G. Requicha, Nat. Mater. **2**, 229-232 (2003).

[16] S. I. Bozhevolnyi, V. S. Volkov, E. Devaux, J. Y. Laluet, and T. W. Ebbesen, Nature **440**, 508-511 (2006).

[17] W. Rechberger, A. Hohenau, A. Leitner, J. R. Krenn, B. Lamprecht, and F. R. Aussenegg, Opt. Commun. **220**, 137-141 (2003).

[18] D. P. Fromm, A. Sundaramurthy, P. J. Schuck, G. Kino, and W. E. Moerner, Nano Lett. **4**, 957-961 (2004).

[19] T. Atay, J. H. Song, and A. V. Nurmikko, Nano Lett. **4**, 1627-1632 (2004).

[20] P. Muhlschlegel, H. J. Eisler, O. J. F. Martin, B. Hecht, and D. W. Pohl, Science **308**, 1607-1609 (2005).

[21] A. Sundaramurthy, K. B. Crozier, G. S. Kino, D. P. Fromm, P. J. Schuck, and W. E. Moerner, Phys. Rev. B **72**, 165409 (2005).

[22] A. K. Sarychev, G. Shvets, and V. M. Shalaev, Phys. Rev. E **73**, 36609 (2006).

[23] A. Sundaramurthy, P. J. Schuck, N. R. Conley, D. P. Fromm, G. S. Kino, and W. E. Moerners, Nano Lett. **6**, 355-360 (2006).

[24] O. L. Muskens, V. Giannini, J. A. Sanchez-Gil, and J. G. Rivas, Nano Lett. **7**, 2871-2875 (2007).

[25] R. M. Bakker, V. P. Drachev, Z. Liu, H. K. Yuan, R. H. Pedersen, A. Boltasseva, J. Chen, J. Irudayaraj, A. V. Kildishev, and V. M. Shalaev, New J. Phys. **10**, 125022 (2008).

[26] R. M. Bakker, H. K. Yuan, Z. Liu, V. P. Drachev, A. V. Kildishev, V. M. Shalaev, R. H. Pedersen, S. Gresillon, and A. Boltasseva, Appl. Phys. Lett. **92**, 043101 (2008).

[27] N. Fang and X. Zhang, Appl. Phys. Lett. **82**, 161-163 (2003).

[28] N. Fang, H. Lee, C. Sun, and X. Zhang, Science **308**, 534-537 (2005).

[29] D. O. S. Melville and R. J. Blaikie, Opt. Exp. **13**, 2127-2134 (2005).

[30] W. Cai, D. A. Genov, and V. M. Shalaev, Phys. Rev. B **72**, 193101 (2005).

[31] D. Schurig, J. J. Mock, B. J. Justice, S. A. Cummer, J. B. Pendry, A. F. Starr, and D. R. Smith, Science **314**, 977-980 (2006).

[32] U. Leonhardt, Science **312**, 1777-1780 (2006).

[33] W. S. Cai, U. K. Chettiar, A. V. Kildishev, and V. M. Shalaev, Nat. Photonics **1**, 224-227 (2007).

[34] A. Alù and N. Engheta, Phys. Rev. Lett. **100**, 113901 (2008).

[35] Z. Jacob, L. V. Alekseyev, and E. Narimanov, Opt. Exp. **14**, 8247-8256 (2006).

[36] A. Salandrino and N. Engheta, Phys. Rev. B **74**, 75103 (2006).

[37] Z. Liu, H. Lee, Y. Xiong, C. Sun, and X. Zhang, Science **315**, 1686-1686 (2007).

[38] I. I. Smolyaninov, Y. J. Hung, and C. C. Davis, Science **315**, 1699 - 1701 (2007).

[39] D. Schurig, J. B. Pendry, and D. R. Smith, Opt. Exp. **15**, 14772-14782 (2007).

[40] Y. Xiong, Z. Liu, and X. Zhang, Appl. Phys. Lett. **94**, 203108 (2009).

[41] P. B. Johnson and R. W. Christy, Phys. Rev. B **6**, 4370-4379 (1972).

[42] J. P. Marton and B. D. Jordan, Phys. Rev. B **15**, 1719-1727 (1977).

[43] A. N. Sudarkin and P. A. Demkovich, Sov. Phys. Tech. Phys **34**, 764-766 (1989).

[44] C. Sirtori, C. Gmachl, F. Capasso, Faist, J, D. L. Sivco, A. L. Hutchinson, and A. Y. Cho, Opt. Lett. **23**, 1366-1368 (1998).

[45] M. P. Nezhad, K. Tetz, and Y. Fainman, Appl. Phys. Lett **77**, 773-775 (2000).

[46] A. Tredicucci, C. Gmachl, F. Capasso, A. L. Hutchinson, D. L. Sivco, and A. Y. Cho, Appl. Phys. Lett. **76**, 2164-2166 (2000).

[47] D. J. Bergman and M. I. Stockman, Phys. Rev. Lett. **90**, 27402 (2003).

[48] N. M. Lawandy, Appl. Phys. Lett. **85**, 5040-5042 (2004).

[49] J. Seidel, S. Grafstroem, and L. Eng, Phys. Rev. Lett. **94**, 177401 (2005).

[50] M. Noginov, G. Zhu, M. Bahoura, J. Adegoke, C. Small, B. Ritzo, V. Drachev, and V. Shalaev, Opt. Lett. **31**, 3022-3024 (2006).

[51] T. A. Klar, A. V. Kildishev, V. P. Drachev, and V. M. Shalaev, IEEE J. Sel. Top. Quantum Electron **12**, 1106-1115 (2006).

[52] M. A. Noginov, V. A. Podolskiy, G. Zhu, M. Mayy, M. Bahoura, J. A. Adegoke, B. A. Ritzo, and K. Reynolds, Opt. Exp. **16**, 1385-1392 (2008).

[53] M. Noginov, G. Zhu, M. Mayy, B. Ritzo, N. Noginova, and V. Podolskiy, Phys. Rev. Lett. **101**, 226806 (2008).

[54] M. Ambati, S. Nam, E. Ulin-Avila, D. Genov, G. Bartal, and X. Zhang, Nano Lett. **8**, 3998-4001 (2008).

[55] N. I. Zheludev, S. L. Prosvirnin, N. Papasimakis, and V. A. Fedotov, Nat. Photonics **2**, 351-354 (2008).

[56] M. A. Noginov, G. Zhu, A. M. Belgrave, R. Bakker, V. M. Shalaev, E. E. Narimanov, S. Stout, E. Herz, T. Suteewong, and U. Wiesner, Nature **460**, 1110-1112 (2009).

[57] R. Oulton, V. Sorger, T. Zentgraf, R. Ma, C. Gladden, L. Dai, G. Bartal, and X. Zhang, Nature **461**, 629-632 (2009).

[58] V. P. Drachev, U. K. Chettiar, A. V. Kildishev, H. K. Yuan, W. Cai, and V. M. Shalaev, Opt. Exp. **16**, 1186-1195 (2008).

[59] N. W. Ashcroft and N. D. Mermin (1st ed), Solid State Physics (Harcourt College Publisher, Philadelphia, 1976), p. 1-20.

[60] H. Hövel, S. Fritz, A. Hilger, U. Kreibig, and M. Vollmer, Phys. Rev. B **48**, 18178-18188 (1993).

[61] B. N. J. Persson, Phys. Rev. B **44**, 3277-3296 (1991).

[62] H. Ehrenreich and H. R. Philipp, Phys. Rev. **128**, 1622-1629 (1962).

[63] B. R. Cooper, H. Ehrenreich, and H. R. Philipp, Phys. Rev **138**, 494–507 (1965).

[64] H. Ehrenreich, H. Philipp, and B. Segall, Phys. Rev. **132**, 1918-1928 (1963).

[65] R. LaVilla and H. Mendlowitz, Phys. Rev. Lett. **9**, 149-150 (1962).

[66] M. Dressel and G. Grüner (1st ed), Electrodynamics of Solids (Cambridge University Press, Cambridge, 2002), p. 136-171.

[67] A. D. Rakic, A. B. Djurisic, J. M. Elazar, and M. L. Majewski, Appl. Opt. **37**, 5271-5283 (1998).

[68] D. G. Thomas, J. J. Hopfield, and C. J. Frosch, Phys. Rev. Lett. **15**, 857-860 (1965).

[69] C. Kittel and P. McEuen (7th ed), Introduction to Solid State Physics (Wiley, New York, 1996), p. 269-324.

[70] J. J. Hopfield, Phys. Rev. **112**, 1555-1567 (1958).

[71] J. D. Dow and D. Redfield, Phys. Rev. B **1**, 3358-3371 (1970).

[72] W. P. Dumke, Phys. Rev. **132**, 1998-2002 (1963).

[73] G. Dolling, M. Wegener, C. M. Soukoulis, and S. Linden, Opt. Lett. **32**, 53-55 (2007).

[74] T. W. Ebbesen, H. J. Lezec, H. F. Ghaemi, T. Thio, and P. A. Wolff, Nature **391**, 667-669 (1998).

[75] V. M. Shalaev, W. Cai, U. K. Chettiar, H. K. Yuan, A. K. Sarychev, V. P. Drachev, and A. V. Kildishev, Opt. Lett. **30**, 3356-3358 (2005).

[76] P. Tobiška, O. Hugon, A. Trouillet, and H. Gagnaire, Sens. Actuators, B **74**, 168-172 (2001).

[77] S. Baldelli, A. S. Eppler, E. Anderson, Y. R. Shen, and G. A. Somorjai, J. Chem. Phys. **113**, 5432-5438 (2000).

[78] E. D. Palik, Handbook of Optical Constants of Solids [electronic resource] (Academic press, San Diego, 1998).

[79] M. G. Blaber, M. D. Arnold, N. Harris, M. J. Ford, and M. B. Cortie, Physica B **394**, 184-187 (2007).

[80] T. Lewowski, Thin Solid Films **259**, 53-58 (1995).

[81] H. Wei and H. Eilers, J. Phys. Chem. Solids **70**, 459-465 (2009).

[82] T. W. H. Oates and A. Mucklich, Nanotechnology **16**, 2606-2611 (2005).

[83] S. K. So, H. H. Fong, and N. H. Cheung, in: Proceeding of the Materials Research Society Symposium O, Vol. 672, Materials Research Society, 2001 Spring, pp.

O10.7.1-O10.7.6.

[84] D. D. Smith, Y. Yoon, R. W. Boyd, J. K. Campbell, L. A. Baker, R. M. Crooks, and M. George, J. Appl. Phys. **86**, 6200 (1999).

[85] Y. Yagil, P. Gadenne, C. Julien, and G. Deutscher, Phys. Rev. B **46**, 2503-2511 (1992).

[86] K. Seal, M. A. Nelson, and Z. C. Ying, J. Mod. Opt. **14**, 2423-2435 (2002).

[87] A. K. Sarychev and V. M. Shalaev (1st ed), Electrodynamics of metamaterials (World Scientific, Singapore, 2007).

[88] G. H. Chan, J. Zhao, E. M. Hicks, G. C. Schatz, and R. P. Van Duyne, Nano Lett **7**, 1947-1952 (2007).

[89] C. Langhammer, M. Schwind, B. Kasemo, and I. Zori Nano Lett. **8**, 1461-1471 (2008).

[90] G. Chan, J. Zhao, G. Schatz, and R. Duyne, J. Phys. Chem. C **112**, 13958-13963 (2008).

[91] J. C. Quail, J. G. Rako, and H. J. Simon, Opt. Lett. **8**, 377-379 (1983).

[92] J. Malicka, I. Gryczynski, Z. Gryczynski, and J. R. Lakowicz, J. Phys. Chem. B **108**, 19114-19118 (2004).

[93] K. Ray, M. H. Chowdhury, and J. R. Lakowicz, Anal. Chem. **79**, 6480-6487 (2007).

[94] T. Dorfer, M. Schmitt, and J. Popp, J. Raman Spectrosc. **38**, 1379-1382 (2007).

[95] A. Taguchi, N. Hayazawa, K. Furusawa, H. Ishitobi, and S. Kawata, J. Raman Spectrosc. **40**, 1324-1330 (2009).

[96] V. Guritanu, A. B. Kuzmenko, D. van der Marel, S. M. Kazakov, N. D. Zhigadlo, and J. Karpinski, Phys. Rev. B **73**, 104509 (2006).

[97] Y. Fudamoto and S. Lee, Phys. Rev. B **68**, 184514 (2003).

[98] R. A. Kaindl, M. A. Carnahan, J. Orenstein, D. S. Chemla, H. M. Christen, H. Y. Zhai, M. Paranthaman, and D. H. Lowndes, Phys. Rev. Lett. **88**, 27003 (2001).

[99] J. J. Tu, G. L. Carr, V. Perebeinos, C. C. Homes, M. Strongin, P. B. Allen, W. N. Kang, E. M. Choi, H. J. Kim, and S. I. Lee, Phys. Rev. Lett. **87**, 277001 (2001).

[100] M. O. Mun, Y. J. Kim, Y. Park, J. H. Kim, S. H. Moon, H. N. Lee, H. G. Kim, and B. Oh, J. Supercond. **15**, 475-477 (2002).

[101] J. T. Lue, S. J. Mu, and I. C. Wu, Phys. Rev. B **36**, 1657-1661 (1987).

[102] A. Borghesi, L. Nosenzo, A. Piaggi, G. Guizzetti, C. Nobili, and G. Ottaviani, Phys. Rev. B **38**, 10937-10940 (1988).

[103] A. Borghesi, A. Piaggi, G. Guizzetti, F. Nava, and M. Bacchetta, Phys. Rev. B **40**, 3249-3253 (1989).

[104] M. Amiotti, A. Borghesi, G. Guizzetti, and F. Nava, Phys. Rev. B **42**, 8939-8946 (1990).

[105] V. N. Antonov, O. Jepsen, O. K. Andersen, A. Borghesi, C. Bosio, F. Marabelli, A. Piaggi, G. Guizzetti, and F. Nava, Phys. Rev. B **44**, 8437-8445 (1991).

[106] Z. C. Wu, E. T. Arakawa, J. R. Jimenez, and L. J. Schowalter, Phys. Rev. B **47**, 4356-4362 (1993).

[107] J. W. Cleary, R. E. Peale, D. Shelton, G. Boreman, R. Soref, and W. R. Buchwald, in: Proceeding of, 2008.

[108] D. A. Bobb, G. Zhu, M. Mayy, A. V. Gavrilenko, P. Mead, V. I. Gavrilenko, and M. A. Noginov, Appl. Phys. Lett. **95**, 151102 (2009).

[109] M. G. Blaber, M. D. Arnold, and M. J. Ford, J. Phys. Condens. Matter **21**, (2009).

[110] M. Zwilling, P. C. Schmidt, and A. Weiss, Appl. Phys. A **16**, 255-269 (1978).

[111] K. H. Kim, K. C. Park, and D. Y. Ma, J. Appl. Phys. **81**, 7764-7772 (1997).

[112] I. Hamberg and C. G. Granqvist, J. Appl. Phys. **60**, 123-159 (1986).

[113] A. J. Hoffman, L. Alekseyev, S. S. Howard, K. J. Franz, D. Wasserman, V. A. Podolskiy, E. E. Narimanov, D. L. Sivco, and C. Gmachl, Nat. Mater. **6**, 946-950 (2007).

[114] J. A. Schuller, R. Zia, T. Taubner, and M. L. Brongersma, Phys. Rev. Lett. **99**, 107401 (2007).

[115] T. Taubner, D. Korobkin, Y. Urzhumov, G. Shvets, and R. Hillenbrand, Science **313**, 1595-1595 (2006).

[116] C. M. Wolfe and G. E. Stillman, Appl. Phys. Lett. **27**, 564-567 (1975).

[117] X. A. Cao, C. R. Abernathy, R. K. Singh, S. J. Pearton, M. Fu, V. Sarvepalli, J. A. Sekhar, J. C. Zolper, D. J. Rieger, and J. Han, Appl. Phys. Lett. **73**, 229-231 (1998).

[118] P. F. Robusto and R. Braunstein, Phys. Status Solidi A-Appl. Res. **119**, 155-168 (1990).

[119] M. Y. C. Xu, M. Z. Alam, A. J. Zilkie, K. Zeaiter, and J. S. Aitchison, in: Proceeding of the Conference on Lasers and Electro-Optics/Quantum Electronics and Laser Science Conference, San Jose, CA, USA, 2008, IEEE, pp. 2135-2136.

[120] F. Michelotti, L. Dominici, E. Descrovi, N. Danz, and F. Menchini, Opt. Lett. **34**, 839-841 (2009).

[121] C. Rhodes, S. Franzen, J. P. Maria, M. Losego, D. N. Leonard, B. Laughlin, G. Duscher, and S. Weibel, J. Appl. Phys. **100**, 054905 (2006).

[122] C. Rhodes, M. Cerruti, A. Efremenko, M. Losego, D. E. Aspnes, J. P. Maria, and S. Franzen, J. Appl. Phys. **103**, 93108 (2008).

[123] S. Franzen, J. Phys. Chem. C **112**, 6027-6032 (2008).

[124] S. H. Brewer and S. Franzen, Chem. Phys. **300**, 285-293 (2004).

[125] S. H. Brewer and S. Franzen, J. Alloys Compd **343**, 244-244 (2002).

[126] A. Solieman and M. A. Aegerter, Thin Solid Films **502**, 205-211 (2006).

[127] K. Ellmer and R. Mientus, Thin Solid Films **516**, 5829-5835 (2007).

[128] F. C. Lai, L. M. Lin, R. Q. Gai, Y. Z. Lin, and Z. G. Huang, Thin Solid Films **515**, 7387-7392 (2007).

[129] M. Hiramatsu, K. Imaeda, N. Horio, and M. Nawata, J. Vac. Sci. Technol. A **16**, 669-673 (1998).

[130] H. Kim, A. Pique, J. S. Horwitz, H. Murata, Z. H. Kafafi, C. M. Gilmore, and D. B. Chrisey, Thin Solid Films **377**, 798-802 (2000).

[131] H. Agura, A. Suzuki, T. Matsushita, T. Aoki, and M. Okuda, Thin Solid Films **445**, 263-267 (2003).

[132] A. V. Singh, R. M. Mehra, N. Buthrath, A. Wakahara, and A. Yoshida, J. Appl. Phys. **90**, 5661-5665 (2001).

[133] J. G. Lu, Z. Z. Ye, Y. J. Zeng, L. P. Zhu, L. Wang, J. Yuan, B. H. Zhao, and Q. L.

Liang, J. Appl. Phys. **100**, 073714 (2006).

[134] M. H. Yoon, S. H. Lee, H. L. Park, H. K. Kim, and M. S. Jang, J. Mater. Sci. Lett. **21**, 1703-1704 (2002).

[135] K. S. Novoselov, A. K. Geim, S. V. Morozov, D. Jiang, Y. Zhang, S. V. Dubonos, I. V. Grigorieva, and A. A. Firsov, Science **306**, 666-669 (2004).

[136] K. S. Novoselov, A. K. Geim, S. V. Morozov, D. Jiang, M. I. Katsnelson, I. V. Grigorieva, S. V. Dubonos, and A. A. Firsov, Nature **438**, 197-200 (2005).

[137] A. K. Geim and K. S. Novoselov, Nat. Mater. **6**, 183-191 (2007).

[138] K. V. Emtsev, A. Bostwick, K. Horn, J. Jobst, G. L. Kellogg, L. Ley, J. L. McChesney, T. Ohta, S. A. Reshanov, J. Rohrl, E. Rotenberg, A. K. Schmid, D. Waldmann, H. B. Weber, and T. Seyller, Nat. Mater. **8**, 203-207 (2009).

[139] F. Stern, Phys. Rev. Lett. **18**, 546-548 (1967).

[140] E. H. Hwang and S. Das Sarma, Phys. Rev. B **75**, 205418 (2007).

[141] S. D. Sarma and E. H. Hwang, Phys. Rev. Lett. **102**, 206412 (2009).

[142] V. Ryzhii, A. Satou, and T. Otsuji, J. Appl. Phys. **101**, 024509 (2007).

[143] F. Rana, IEEE Trans. Nanotechnol. **7**, 91-99 (2008).

[144] A. A. Dubinov, V. Y. Aleshkin, M. Ryzhii, T. Otsuji, and V. Ryzhii, Appl. Phys. Exp. **2**, 3 (2009).

[145] M. Jablan, H. Buljan, and M. Soljacic, Arxiv preprint arXiv:0910.2549 (2009).

[146] Z. Q. Li, E. A. Henriksen, Z. Jiang, Z. Hao, M. C. Martin, P. Kim, H. L. Stormer, and D. N. Basov, Nat. Phys. **4**, 532-535 (2008).

[147] J. D. Jackson (3rd ed), Classical Electrodynamics (Wiley, New York, 1999), p. 371-374.

[148] V. M. Shalaev, in: Proceeding of the International School on Quantum Electronics, Erice, Sicily, Italy, 2-14 July 2000, American Institute of Physics, pp. 239-243.

[149] F. Wang and Y. R. Shen, Phys. Rev. Lett. **97**, 206806 (2006).

[150] P. Berini, Opt. Exp. **14**, 13030-13042 (2006).

[151] M. D. Arnold and M. G. Blaber, Opt. Exp. **17**, 3835-3847 (2009).

[152] H. Raether (1st ed), Surface plasmons on smooth and rough surfaces and on gratings (Springer-Verlag, Berlin, 1988), p. 5.

[153] S. A. Ramakrishna, J. B. Pendry, M. C. K. Wiltshire, and W. J. Stewart, J. Mod. Opt. **50**, 1419-1430 (2003).